\documentclass[%
 reprint,
superscriptaddress,
 amsmath,amssymb,
 aps,
 pra
]{revtex4-2}

\usepackage{graphicx}% Include figure files
\usepackage{dcolumn}% Align table columns on decimal point
\usepackage{bm}% bold math
\usepackage{soul}
\usepackage[colorlinks]{hyperref}

\usepackage[english]{babel}
\usepackage{color}
\usepackage{mathtools}
\usepackage{cleveref}
\usepackage{verbatim}
\usepackage{breqn}

\makeatletter
\newcommand{\customlabel}[2]{%
   \protected@write \@auxout {}{\string \newlabel {#1}{{#2}{\thepage}{#2}{#1}{}} }%
   \hypertarget{#1}{}
}
\makeatother

\begin{document}

%\title{Excitability in the asymmetrically-driven dissipative photonic Bose-Hubbard dimer}

\title{Neuron-like spiking dynamics in the asymmetrically-driven dissipative photonic Bose-Hubbard dimer}
\author{Jes\'us Yelo-Sarri\'on}

\author{Francois Leo}
\author{Simon-Pierre Gorza}
\affiliation{OPERA-Photonique$,$ Université libre de Bruxelles$,$ 50 Avenue F. D. Roosevelt$,$ CP 194/5 B-1050 Bruxelles$,$ Belgium}

\author{Pedro Parra-Rivas}
\email{pedro.parra-rivas@uniroma1.it}
\affiliation{OPERA-Photonique$,$ Université libre de Bruxelles$,$ 50 Avenue F. D. Roosevelt$,$ CP 194/5 B-1050 Bruxelles$,$ Belgium}

\affiliation{Dipartimento di Ingegneria dell’Informazione$,$ Elettronica e Telecomunicazioni$,$
Sapienza Universit\'a di Roma$,$ via Eudossiana 18$,$ 00184 Rome$,$ Italy\\}

\date{\today}

\begin{abstract}
We demonstrate neuron-like spiking dynamics in the asymmetrically driven dissipative photonic Bose-Hubbard dimmer model which describes two coupled nonlinear passive Kerr cavities. Spiking dynamics appear due to the excitable nature of the system. In this context, excitable excursions in the phase space correspond to spikes in the temporal evolution of the field variables. In our case, excitability is mediated by the destruction of an oscillatory state in a global homoclinic bifurcation. In this type of excitability (known as type-I) the period of the oscillatory state diverges when approaching the bifurcation. Beyond this point, the system exhibits excitable dynamics under the application of suitable perturbations. We have also characterized the effect that additive Gaussian noise has on the spiking dynamics, showing that the system undergoes a coherence resonance for a given value of the noise strength.
\end{abstract}

\maketitle
    
\section{Introduction}
\begin{figure}
\includegraphics[width=\columnwidth]{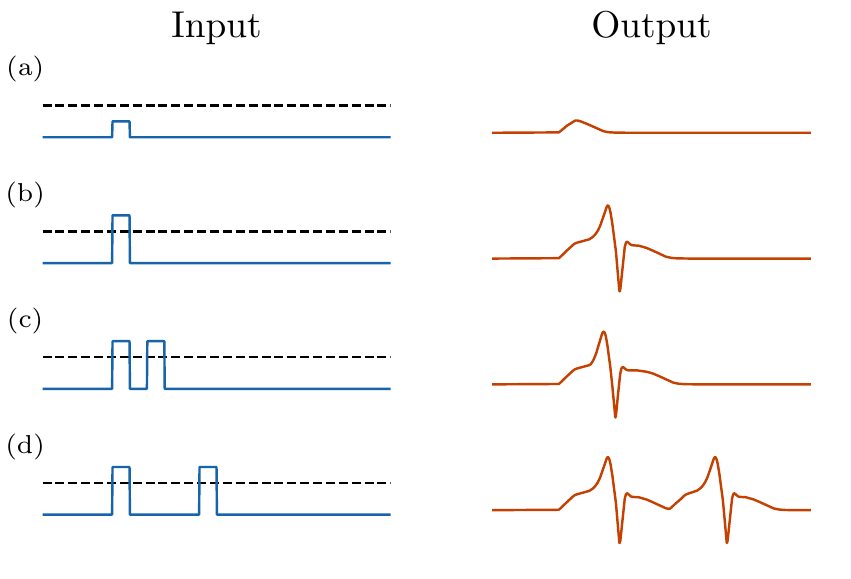}
\hspace{1cm}
\includegraphics[width=\columnwidth]{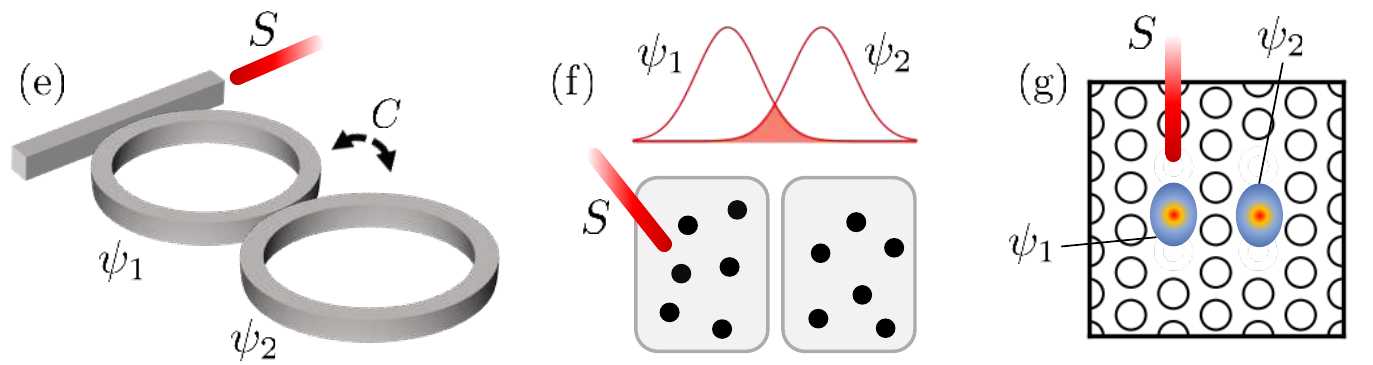}
\caption{Panels (a)-(d) show how different kinds of inputs (left column) cause different kinds of responses (right column). An input below threshold (dashed line) induces a fast decaying to the resting state [see (a)]; (b) an input above threshold trigger a spike; (c) two impulses generated within the refractory time yield a single spike, however for time separations larger than the refractory time two spikes can be generated [see (d)].  Panel (e) shows a schematic representation of an asymmetrically-driven dissipative photonic Bose-Hubbard dimer, (f) corresponds to a schematic of polaritons in nanopillars and (g) to a photonic crystal cavity. }
\label{fig1}
\end{figure}

Spiking-like dynamics were observed for the first time in biology in the context of neuronal and heart cells \cite{hodgkin,kraepelin__1981,koch_biophysics_2004}. Spiking dynamics is related to the concept of {\it excitability}. The main features of an excitable system are related to how the system responds to external perturbations from a stable steady resting state. The nature of this response depends on the amplitude of the perturbation compared to a certain threshold. For perturbations below the threshold, the system decays very fast to the resting state [see Fig.~\ref{fig1}(a)], while for perturbations above such threshold the system experiences a nontrivial trajectory (excursion) in the phase space before returning to the resting state [see Fig.~\ref{fig1}(b)]. This large excursion is independent of the characteristics of the perturbation and corresponds to the so-called {\it spike} observed in the temporal trace. One essential feature of excitability is that once a spike is triggered, the system is unable to initiate another excursion during the recovery time (or refractory period). 
%needs a recovery (or refractory) time to excite a second excursion 
[see Figs.~\ref{fig1}(c),(d)].
This dynamical process is what makes neurons able to perform computations and process externals inputs \cite{kraepelin__1981,koch_biophysics_2004,izhikevich_neural_2012}. 

In general, excitability requires the destruction of a permanent oscillatory state (i.e., a limit cycle) of finite-amplitude as a suitable parameter of the system is modified. 
%modifying a suitable parameter of the system. 
Thus, an excitable excursion would follow the remnants of the cycle before returning to the resting state \cite{kraepelin__1981,izhikevich_neural_2012}. 

Since its discovery, excitability has been observed in a wide range of natural systems, which include, chemical reactions \cite{neumann_physical_1977,kuhnert_image_1989}, biology \cite{murray_mathematical_2007,shepherd_neurobiology_1994,braun_oscillation_1994,adamatzky_spiking_2018}, Josephson junctions \cite{Arindam}, and laser physics\cite{weis_dynamics_1991}, to only cite a few. In the latter, some examples include  semiconductor lasers \cite{wunsche_excitability_2001,beri_excitability_2010,gelens_excitability_2010,garbin_incoherent_2014},  lasers with saturable absorbers \cite{dubbeldam_excitability_1999,larotonda_experimental_2002}, lasers with optical feedback \cite{giudici_andronov_1997} silicon-on-insulator microrings \cite{VanVaerenbergh:12,vaerenbergh_cascadable_2012,van_vaerenbergh_simplified_2012,vaerenbergh_towards_nodate}, and microdisk lasers \cite{alexander_excitability_2013}. The main goal of these studies is to propose an effective optical spiking neuron design on a chip. 

\customlabel{fig:1}{1}
\customlabel{fig:1a}{1(a)}
\customlabel{fig:1b}{1(b)}
\customlabel{fig:1c}{1(c)}
\customlabel{fig:1d}{1(d)}
\customlabel{fig:1e}{1(e)}
\customlabel{fig:1f}{1(f)}
\customlabel{fig:1g}{1(g)}

Excitability may also arise in systems with some extended dimension, i.e. in the presence of spatial coupling. In this context, excitable waves can emerge in extended systems which are locally excitable \cite{meron_pattern_1992,murray_mathematical_2007,mikhailov_foundations_1994}. However, the spatial coupling can be responsible for the coherent structures emerging from the spiking dynamics even in systems that are non locally excitable  \cite{gomila_excitability_2005,parra-rivas_dissipative_2013}, and for excitable-like behaviors which stem from front interactions \cite{parra-rivas_front_2017}. Recent works have also focused on the characterization of travelling pulses in type-I excitable media \cite{pablo1, pablo2}.

In this paper, we demonstrate, for the first time to our knowledge, the emergence of spiking dynamics in the context of two coupled passive cavities like those shown schematically in Figs.~\ref{fig1}(e-g). In what follows, we refer to these systems as driven dissipative photonic Bose-Hubbard (DDBH) dimer \cite{Abbarchi2013}, in analogy with the open quantum boson system \cite{Bruder2005}. Furthermore, this model can also describe the dynamics between interacting
bosons for open quantum systems \cite{Bruder2005}. In its simplest realization, two macroscopic phase coherent wave functions
are coupled to form a Bose-Hubbard dimer, which can be also referred to  as a bosonic Josephson junction \cite{Vijay, Arindam}.

Here we show that the asymmetrically driven dissipative Bose-Hubbard dimer possesses all the ingredients required for the emergence of excitable behavior: a stable steady resting state and the destruction of a limit cycle occurring close to it. To do so, we perform a detailed bifurcation analysis which allows us to identify global bifurcations which are responsible for this kind of dynamics. 

The paper is organized as follows. First, we will shortly introduce the mean-field model that we study (see Sec.~\ref{sec:1}). After that, we present the bifurcation structure of the system and identify the most relevant bifurcations (see Sec.~\ref{sec:2}).  Section~\ref{sec:3} is devoted to the characterization of the excitable dynamics, and in Sec.~\ref{sec:4} we show how noise can affect such type of behavior. Finally, in Sec.~\ref{sec:5} we draw some final conclusions. 

\section{Asymmetrically-driven dissipative Bose-Hubbard model}\label{sec:1}

In the absence of dispersion, photonic dimers, like the one shown in Fig.~\ref{fig:1e}, can be described by the mean-field model 
\begin{equation}
\begin{array}{lll}
    \displaystyle \frac{d \psi_1}{dt} &=&\displaystyle [-1 +i(|\psi_1|^2 - \Delta_1)]\psi_1 + iC\psi_2+S\\
    \\
    \displaystyle \frac{d \psi_2}{dt} &=& \displaystyle[-1 +i(|\psi_2|^2 - \Delta_2)]\psi_2  + iC\psi_1,
\label{eq0}
\end{array}
\end{equation}
%where $t=t'\kappa/T_R$ with $t'$ being the laboratory time, $T_R$ the round-trip time and $\kappa$ the cavity loss coefficient (excluding the middle coupler). The detuning from the closest (single cavity) resonances is $\Delta=\delta/\kappa=(m2\pi-\varphi)/\kappa$, where $\varphi$ is the round-trip linear phase shift and $m$ an integer number. The coupling parameter between the two cavities is $C=\sqrt{\theta_{12}}/\kappa$, where $\theta_{12}$ is the transmission coefficient of the coupler between the cavities. The normalized complex fields read $\psi_j=A_j\sqrt{\gamma L / \kappa}$, ($j=1,~2$), where $A_j$ are the field amplitudes normalized such that the intracavity powers (expressed in watts) are given by $|A_j|^2=P_j$. $S=i\sqrt{P_p\gamma L \theta_p/ \kappa^3}$, where $P_p$ is the driving power, $\theta_p$ the transmission coefficient of the input coupler, and $\gamma$ the nonlinear parameter of the waveguide and $L$ is the length of the resonators. 
where $t$ is the time, $\Delta_i$ the detuning from the closest (single cavity) resonance, $C$ the coupling parameter between the two cavities,  $\psi_j$ ($j=1,~2$)  the complex fields and $S$ the driving power. These parameters and variables are dimensionless.

\begin{figure}
\customlabel{fig:2}{2}
\customlabel{fig:2a}{2(a)}
\customlabel{fig:2b}{2(b)}
\customlabel{fig:2c}{2(c)}
\includegraphics[width=\columnwidth]{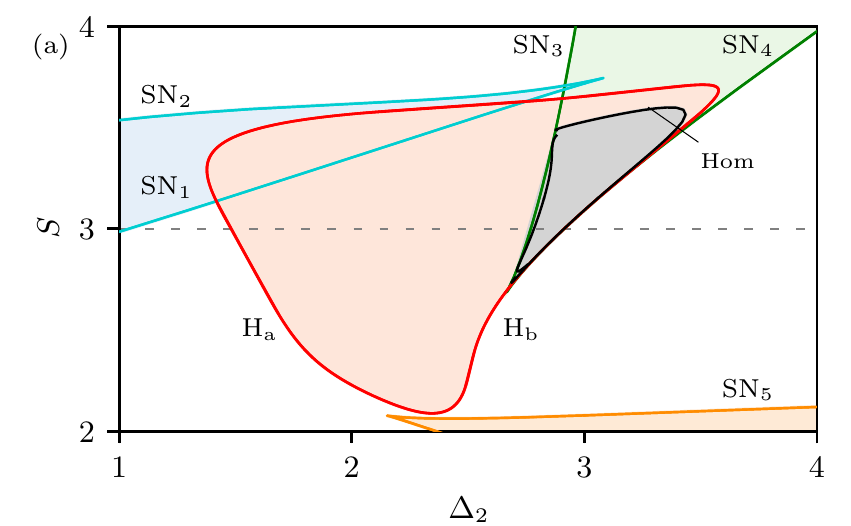}
\includegraphics[width=\columnwidth]{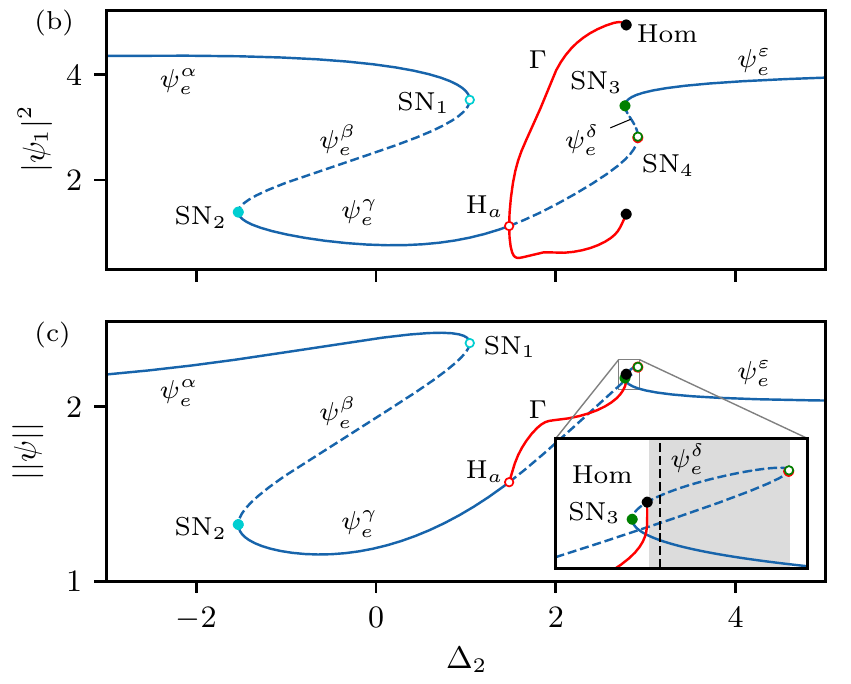}
\caption{(a) Phase diagram in the ($\Delta_2,~S$) parameter space for $\Delta_1=3.1$. The SN$_j$ lines refer to saddle-node bifurcations, bistability occurs inside these regions. The Hopf bifurcations H$_{a,b}$ (red lines) mark the boundaries of the self-pulsing region and the Hom bifurcation (black line) marks the boundaries of the area where there is no self-pulsing. (b) Normalized intracavity power $|\psi_{1}|$ as a function of $\Delta_2$ at a driving value of S = 3.0. Red lines show the maximum and minimum oscillation amplitudes. Note that H$_b$ is located at SN$_4$. (c) $L_2$-norm as a function of $\Delta_2$. The inset shows a close-up view of the region where excitability may occur (gray). The dashed vertical ($\Delta_2=2.785$) line corresponds to the value where Fig. \ref{fig4} is done.}  
\label{fig2}
\end{figure}
In a previous work, we have analyzed the temporal dynamics arising in this system \cite{yelo-sarrion_self-pulsing_2021,sarrion_self-pulsing_2022}. Applying bifurcation analysis and path-continuation techniques \cite{DoedelI,DoedelII}, through the free software package AUTO-07p \cite{Doedel2009}, we were able to classify the region of existence of the different dynamical regimes such as self-pulsing oscillations and chaos \cite{sarrion_self-pulsing_2022}. This analysis predicted that for a coupling constant value $C\approx 1$ the continuous-wave state becomes unstable for $\Delta_1\gtrsim2.1$, leading to stable permanent oscillations. As oscillatory dynamics is needed for excitability, in what follows we fix $C = 1.1$ and $\Delta_1=3.1$.

\section{Phase diagram and temporal dynamics}\label{sec:2}

Fig.~\ref{fig:2a} shows the dynamical regions in the $(\Delta_2,S)$-phase diagram \cite{yelo-sarrion_self-pulsing_2021}. The red shadowed region bounded by the Hopf bifurcation H corresponds to self-pulsing oscillations. The blue and green regions show the coexistence of different homogeneous states which are the equilibrium points of Eq.~(\ref{eq0}). 

To better understand the dynamical organization of the system it is useful to slice Fig.~\ref{fig:2a} by fixing either $\Delta_2$ or $S$.
Figure~\ref{fig:2b} shows one such slice for $S=3.1$, where the intensity of the homogeneous state in the first cavity, $|\psi_1|^2$, is plotted as a function of $\Delta_2$. This curve corresponds to the nonlinear resonance of the cavity. For this value of $\Delta_2$, H is intersected at two different points that we label H$_{a,b}$. 

On the left of the resonance, the stable equilibria $\psi_e^\alpha$ 
and $\psi_e^\gamma$ coexist in-between SN$_1$ and SN$_2$, and are linked through the unstable state $\psi_e^\beta$. Increasing $\Delta_2$, $\psi_e^\gamma$ encounters H$_a$, losing stability in favor of autonomous oscillations. In dynamical systems terms, this state is known as a limit cycle, and we label it $\Gamma$. $\Gamma$ arises from H$_a$ supercritically [see red curve in Fig.~\ref{fig:2b}], and the amplitude of the oscillations increases with $\Delta_2$. Eventually, $\Gamma$ dies at a global homoclinic bifurcation Hom$_a$. This bifurcation corresponds to the Hom black line plotted in Fig.~\ref{fig:2a}. Thus, in the gray region bounded by this line, self-sustained oscillations are absent.

It is easier to see how the cycle touches the unstable equilibrium $\psi_e^\delta$ by plotting the $L2-$norm 
\begin{equation}
||\psi||\equiv \sqrt{T^{-1}\int_0^T(|\psi_1(t)|^2+|\psi_2(t)|^2)dt},
\end{equation}
as a function of $\Delta_2$, where $T$ is the period of the oscillatory state. With this visualization, we can easily see how the cycle disappears at Hom. Above this point, the system falls into the stable homogeneous equilibrium $\psi_e^\epsilon$.
From the right of the resonance, periodic oscillations also emerge from H$_b$, but soon after that, they die at Hom. 

\begin{figure}
\customlabel{fig:3}{3}
\customlabel{fig:3a}{3(a)}
\customlabel{fig:3i}{3(i)}
\customlabel{fig:3ii}{3(ii)}
\customlabel{fig:3iii}{3(iii)}
\customlabel{fig:3iv}{3(iv)}
\includegraphics[scale=0.9]{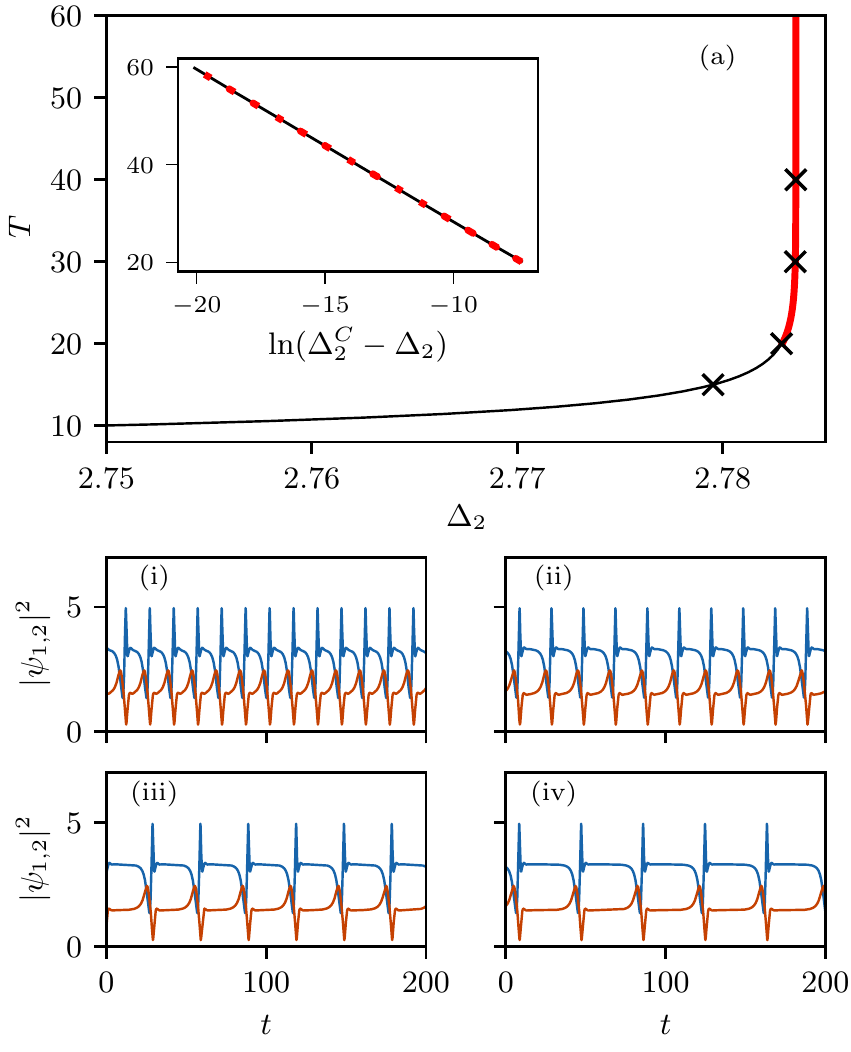}
\caption{(a) Divergence of the cycle's period $T$ as a function of $\Delta_2$ when approaching $\mathrm{Hom}_a$. The close-up view shows the linear dependence of $T$ as a function of ${\rm ln}(\Delta_2^c-\Delta_2)$ (solid line). Dotted line is obtained through a linear fit. The crosses show where simulations of plots (i), (ii), (iii) and (iv) are performed. These oscillations correspond respectively to periods $T~=~15,~20,~30$ and $40$. Here,  $\Delta_1= 3.1$, $C= 1.1$ and $S= 3$.
}  
\label{fig3}
\end{figure}

This homoclinic bifurcation involves the collision of a limit cycle and a saddle-node type equilibrium and is known as tame Shilnikov homoclinic bifurcation \cite{homburg_homoclinic_2010,sarrion_self-pulsing_2022}. These kind of bifurcations are characterized by the exponential divergence of the cycles period $T$ when approaching Hom. This behavior is depicted in Fig.~\ref{fig:3a}, here the divergence (see red segment) follows the scaling law given by 
\begin{equation}
T \propto -\lambda_u^{-1}{\rm ln}(\Delta_2^c-\Delta_2),   
\end{equation}
 where $\lambda_u>0$ is the (unstable) real eigenvalue associated with $\psi_e^\delta$, and $\Delta_2^c$ is the value where Hom occurs. The fit shown in inset of Fig.~\ref{fig:3a} leads a slope of 3.11 that is in close agreement with the theoretical value of $1/\lambda_u = 3.12$. The evolution of $T$ is illustrated in Figs.~\ref{fig3}(i)-(iv) for $T~=~15,~20,~30$ and $40$ (see crosses in Fig~\ref{fig:3a}).

%of the oscillatory state just  
%A particular feature of this homoclinic bifurcation of this type is that the period of the oscillations $T$, existing  diverges exponentially when approaching it and follows a characteristic  Fig.~\ref{fig:3a} shows the divergence of the period as a function of $\Delta_2$ close to the homoclinic bifurcation.
%The red segment is where the linear fit shown in the close up view has been done.  

%When approaching the homoclinic bifurcation, the oscillations turn into a periodic spiking. Numerical simulations corresponding to periods $T~=~15,~20,~30$ and $40$ (crosses in Fig~\ref{fig:3a}) are shown in Figs.~\ref{fig:3i}-\ref{fig:3iv} respectively.
%----------------------------------------------------------------------------------

 %At around $\Delta_2=1.5$, at HB$_1$, self starting oscillations occur until they die at the Homoclinic bifurcation Hom$_1$. Note that the oscillations borned in the right side, at HB$_2$ die directly in the Homoclinic bifurcation Hom$_2$. As a result, SN$_4$, HB$_2$ and Hom$_2$ are overlapped.

%. At a certain critical value a global bifurcation
%takes place: the cycle touches the lower branch LS and
%becomes a homoclinic orbit [Fig. 2(c)]. This is an infiniteperiod bifurcation called saddle loop or homoclinic bifurcation [19]

%Such a situation is depicted in Fig.~\ref{fig1} for $\Delta_L=3.1$ where the bistable regions as well as the self-pulsing area in the ($\Delta_R,\,S_L$)-plane of the parameter space are plotted.
%%%%%%%%%%%%%%%%%
%%%%%%%%%%%%%%%%%   FIG 2

%In order to reduce the dimensionality of the system, we re-write the system as a function of the pseudo-Stokes parameters.

%\begin{equation}
%\begin{array}{lll}
 %   \displaystyle S_1 &=&|\psi_1|^2-|\psi_2|^2\\
  %  \\
   %  \displaystyle S_2 &=& 2|\psi_1\psi_2|\cos(\phi_2-\phi_1)\\
    % \\
%\displaystyle S_3 &=&- 2|\psi_1\psi_2|\sin(\phi_2-\phi_1)

%\label{eq-coupled-lle-diff}
%\end{array}
%\end{equation}
\section{Excitability}\label{sec:3}

The presence of homoclinic bifurcations [see Fig.~\ref{fig:2c}] may lead to excitability of type-I and spiking dynamics \cite{izhikevich_neural_2012}. Here, the stable fixed point (resting state) is $\psi_e^\epsilon$ while $\psi_e^\delta$ plays the role of the perturbation threshold. The region where excitability may occur is delimited by the Hom and the SN$_4$ bifurcations (see gray region in inset).

Figure~\ref{fig:4} illustrates this dynamical behaviour. In Fig.~\ref{fig:2a} we show the ($|\psi_1|^2,|\psi_2|^2$)-phase plane
where the three equilibria $\psi_e^{\gamma,\delta,\epsilon}$ are depicted.
The nature of these points is determined by their associated eigenvalues $\{\lambda_i\}$, which are plotted in Fig.~\ref{fig5}. Thus, $\psi_e^{\epsilon}$ is a 1-spiral~sink, and therefore stable, [see Fig.~\ref{fig:5i}], $\psi_e^{\delta}$ is a spiral-3:1~saddle [see Fig.~\ref{fig:5ii})], and  $\psi_e^{\gamma}$ is a 2:spiral-2~saddle [see Fig.~\ref{fig:5iii}] \cite{hofmann_visualization_2018}. The two unstable fixed points are different types of saddle-focus equilibria \cite{wiggins_introduction_2003}.
\begin{figure}
\customlabel{fig:4}{4}
\customlabel{fig:4a}{4(a)}
\customlabel{fig:4b}{4(b)}
\customlabel{fig:4c}{4(c)}
\includegraphics[width=\columnwidth]{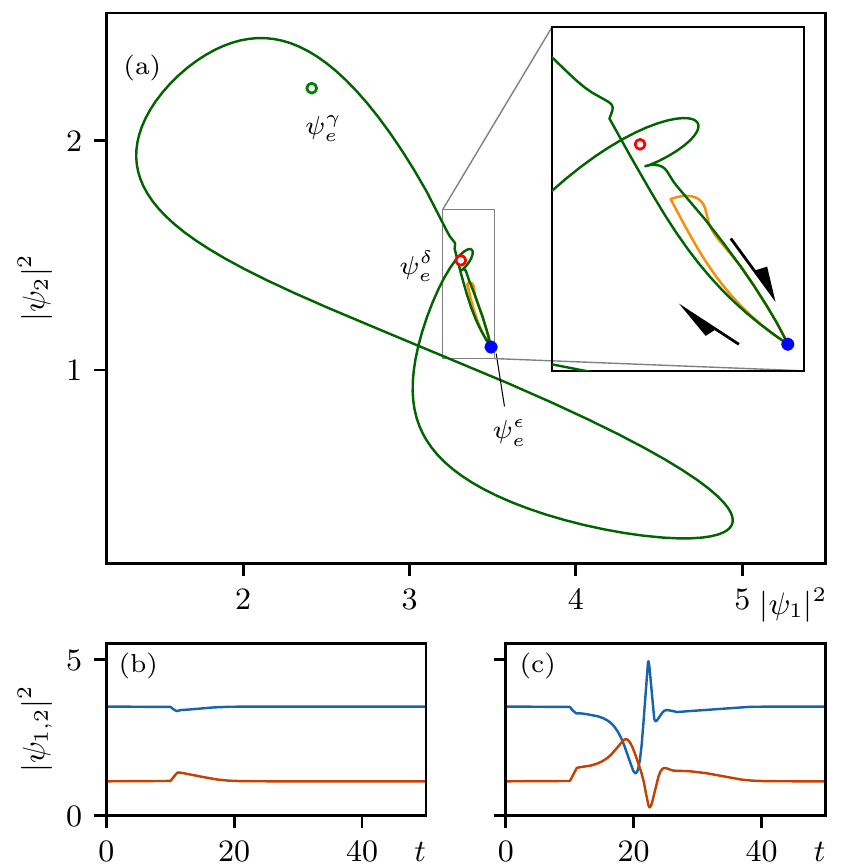}
\caption{Spike excursions following a pulse in the direction of $\eta\hspace{0.3mm}\vec{r_u}$ for $\Delta_2=2.785$ and $S=3.0$.
%and $D=1e-4$.
 (a)  ($|\psi_1|^2,~|\psi_2|^2$)-phase space. We depict fixed points with one or two repelling eigenvalues with a circle and fully attracting fixed points with a dot. In the close-up view, we show the perturbation and the phase spaces of the (b) unsuccessful (orange solid line, $\eta =2$) and (c) successful (green solid line, $\eta = 3$) spike excitation.}  
\label{fig4}
\end{figure}

\begin{figure}
\customlabel{fig:5i}{5(i)}
\customlabel{fig:5ii}{5(ii)}
\customlabel{fig:5iii}{5(iii)}
\customlabel{fig:4}{4}
%\customlabel{fig:4a}{4(a)}
%\customlabel{fig:4b}{4(b)}
%\customlabel{fig:4c}{4(c)}
\includegraphics[width=\columnwidth]{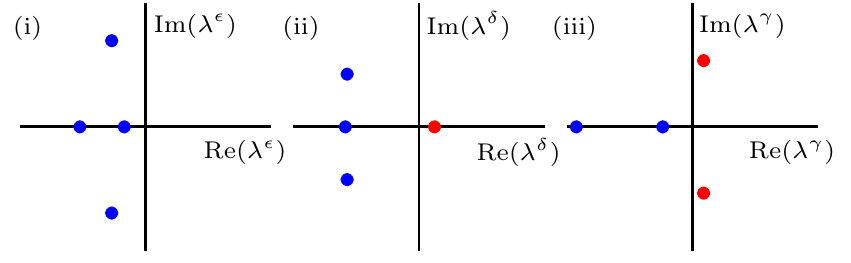}
\caption{Classification of critical point types of the 4D vector field (i) 1-spiral~sink, (ii) spiral-3:1~saddle and (iii) 2:spiral-2~saddle corresponding to $\psi_e^\epsilon$, $\psi_e^\delta$, and $\psi_e^\gamma$, respectively (see blue, red and green dots in Fig.~\ref{fig:4a}, respectively).}
\label{fig5}
\end{figure}

\begin{figure*}
\customlabel{fig:6a}{6(a)}
\customlabel{fig:6b}{6(b)}
\customlabel{fig:6c}{6(c)}
\includegraphics[width=\textwidth]{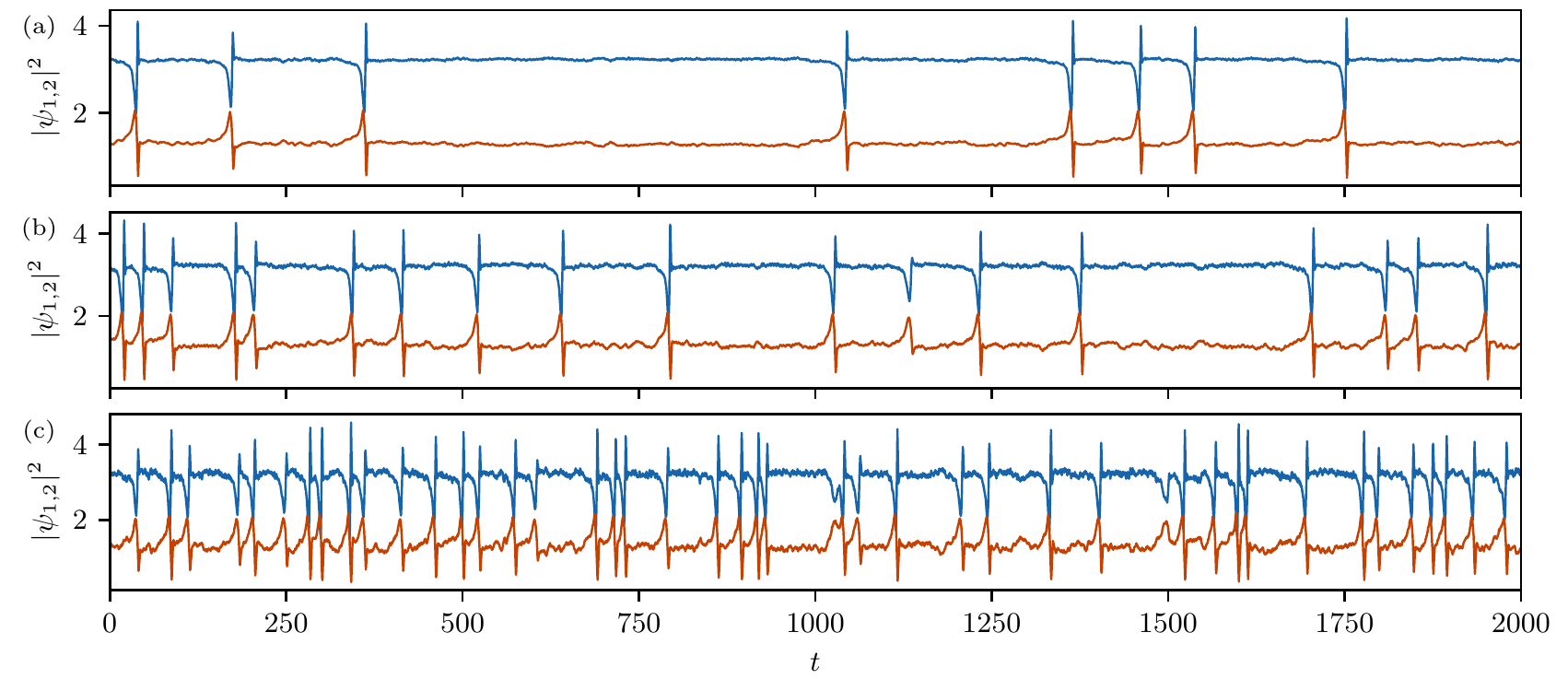}
\caption{
Stochastic dynamics in the DDBH model (\ref{eq_stoch}) showing random spiking behaviour and coherence resonance. The numerical simulations are performed for $S=2.8$, $\Delta_2 = 2.7112$ and different values of noise amplitude $\sqrt{D}$. In (a) $\sqrt{D} = 0.005$, in (b) $\sqrt{D}=0.01$, and in (c) $\sqrt{D}=0.02$. These values correspond to the coherence resonance curve plotted in Fig.~\ref{fig7}.}  
\label{fig6}
\end{figure*}

\begin{figure}
\includegraphics[width=\columnwidth]{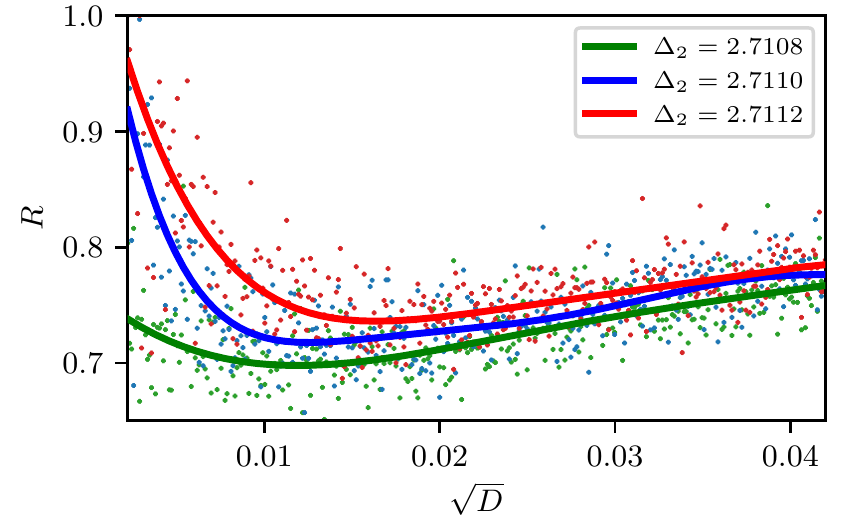}
\caption{%Random spiking behaviour due to the effect of white Gaussian noise on the excitable regime. Excitable excursions with Gaussian white noise 
Coherence resonance indicator $R$ as a function of $\sqrt{D}$ for $S=2.8$ and $\Delta_2 = 2.7108,~2.7110$ and $2.7112$ (green, blue and red). Dots correspond to numerical simulations and lines to $6^{th}$ order polynomial fit.}  
\label{fig7}
\end{figure}

In order to excite the system and trigger a spike, we apply to $\psi_e^\epsilon$ the perturbation $\eta\hspace{0.3mm}\vec{r_u}$,
where $\vec{r_u}$ corresponds to the unstable eigenvector of $\psi_e^\delta$. For $\eta=2$ 
[see Fig.~\ref{fig:4b}] the perturbation is not large enough to cross the threshold, and no excitable excursion takes place. This excursion corresponds to the orange trajectory shown in the ($|\psi_1|^2,~|\psi_2|^2$)-phase plane of Fig.~\ref{fig:4a}. 
Increasing the amplitude of the perturbation to $\eta=3$ is enough to overpass the threshold and excite the system, leading to the spike shown in Fig.~\ref{fig:4c}. This leads to a long transient in the phase space, i.e., an excitable excursion [see green curve in Fig.~\ref{fig:4a}], before returning to the rest state $\psi_e^\epsilon$. The excitable trajectory follows the remnants of the limit cycle before ultimately reaching the resting state. 
%When crossing the threshold $\psi_e^\delta$, the perturbation falls into the trajectory tangent to the eigenvector associated to the unstable eigenvalue of $\psi_e^\delta$.

In the $(\Delta_2,~S)$-phase diagram, the system is excitable all along the homoclinic bifurcation (see Hom in Fig.~\ref{fig:2a}). The closer $\psi_e^{\epsilon,\delta}$ are from Hom, the lower the perturbation amplitude needed to trigger the excitation.

The encountered excitability is of type-I as it is characterized by the divergence of the period of the oscillations occurring close to the bifurcation where is destroyed and for the presence of a real perturbation threshold. In contrast, type-II excitability (not found here) is mediated by Hopf bifurcations, and a well defined threshold does not exist. This is the type of excitability arising in the Fitzhugh-Nagumo models \cite{izhikevich_neural_2012}.

\section{Effect of noise and coherent resonance}\label{sec:4}

The presence of noise in any real experimental setup is unavoidable, and it could have interesting implications in the context of excitability. The interplay between noise and excitability has been studied in different contexts \cite{lindner_effects_2004,jacobo_effects_2010,parra-rivas_front_2017} and may randomly trigger excitable excursions, even for subthreshold perturbations. The emergence of oscillations due to the presence of noise is known as {\it coherence resonance} or internal stochastic resonance \cite{pikovsky_coherence_1997,lindner_effects_2004,jacobo_effects_2010}, and has been analyzed in detail in the context of noise-driven excitable systems \cite{lindner_effects_2004}. 

The dynamics of our system in the presence of stochastic terms is described by the Langevin type equations \cite{Zambon2020}
\begin{equation}
\begin{array}{lll}
    \displaystyle d \psi_{1,2}(t) &=&\displaystyle \left(\kappa_{1,2}(t) + iC\psi_{2,1}(t)\right)dt + d\xi_{1,2}(t)\\
    \\
    \displaystyle \kappa_{j}(t) &=& \displaystyle[-1 +i(|\psi_{j}(t)|^2 - \Delta_{j})]\psi_{j}(t) + S_j
\label{eq_stoch}
\end{array}
\end{equation}
where $\xi_j=\xi_j(t),~j=1,2$, and $S_2=0$ since only one cavity is driven. 

To illustrate the impact of noise on the system we consider an additive complex Gaussian white noise of amplitude $\sqrt{D}$, satisfying
\begin{equation}
\begin{array}{rcl}
    \displaystyle \langle\xi_j(t)\rangle&=&0,\\
    \\
    \displaystyle \langle\xi_j^*(t_a)\xi_{j'}(t_b)\rangle&=&\delta_{j,j'}\delta(t_a-t_b)D.
\label{eq_gauss}
\end{array}
\end{equation}
To solve the stochastic system (\ref{eq_stoch}), we integrate it applying a Heun method \cite{toral_stochastic_2014}.  %applying the  Adams-Bashford-Multon predictor-corrector fourth-order method, warmed up by a lower order Heun method 

The numerical simulation leads to the stochastic dynamics shown in Fig.~\ref{fig:6a}, for $\Delta_2 = 2.7112$ and a noise amplitude $\sqrt{D}=0.005$, where we plot the intensities of the field in both cavities ($|\psi_{1,2}|^2$) as a function of time. As can be seen in Fig.~\ref{fig:6a},  this noise level is able to randomly excite the system, triggering the spiking behavior. However, spikes are rarely triggered, and the intervals between spikes, that hereafter we refer to as $\tau$, are very long and irregular.
Increasing a bit $\sqrt{D}$, i.e., for a moderate noise [see  Fig.~\ref{fig:6b}] the spiking is more regular. This means that, in general, the interspike time interval $\tau$ does not differ that much. For even larger values of $\sqrt{D}$ [see  Fig.~\ref{fig:6c}], the spiking is more frequent. Above a certain value of $\sqrt{D}$, the frequency of spiking continues increasing, although the time between spikes is more irregular again (not shown here).

The variability of the inter-spiking time $\tau$, and therefore the coherence resonance itself, can be quantified using the {\it coefficient of variation}
\begin{equation}
    R=\sqrt{\langle \Delta \tau^2\rangle}/\langle \tau\rangle,
\end{equation}
where $\langle \tau\rangle$ is the mean and $\langle \Delta \tau^2\rangle$ is its variance \cite{pikovsky_coherence_1997,lindner_effects_2004,jacobo_effects_2010}. Periodic spiking is characterized by $R=0$, and a Poisson process (i.e., random spiking) has $R=1$.
The value of $R$ goes through a minimum for intermediate values of noise, indicating that spiking takes place on a more regular basis. The existence of a minimum indicates the occurrence of the coherence resonance \cite{lindner_effects_2004}. Figure~\ref{fig7} shows the coefficient of variation as a function of the noise intensity $\sqrt{D}$ for different values of $\Delta_2$. The red curve ($\Delta_2 = 2.7112$) corresponds to the time traces depicted in Fig.~\ref{fig6}, the spiking dynamics corresponding to its minimum is shown in Fig.~\ref{fig:6c}. For other values of $\Delta_2$, i.e., for different separations from the Hom bifurcation, the coherence resonance indicator behaves similarly. We have verified that when the noise is directly added to the detuning through the term $i\psi_j(t)\xi_j(t)$ the coherence resonance is similar to the one showed in Fig.~\ref{fig7}.

This phenomenon differs from the stochastic resonance, where the characteristic time scale comes from external driving. Here, the noise activates a hidden characteristic time scale (time need to excite the system plus the refractory time) of the system due to its excitable nature.

\section{Conclusions}\label{sec:5}
We have characterized the neuron-like spiking behavior emerging in asymmetrically-driven dissipative Bose-Hubbard model describing photonics dimers in the form of two coupled nonlinear Kerr passive cavities [see Sec.~\ref{sec:1}]. Spiking dynamics appear because the system is excitable, i.e., an external perturbation may cause a different response on the system depending on the amplitude of the perturbation. In this context, a spike corresponds to a long transient response of the system once a certain threshold is overpassed. This long response is commonly known as an excitable excursion. Excitable behavior is related to the appearance/destruction of a limit cycle (i.e., periodic oscillations) when varying a suitable parameter. 

Applying well-known results of dynamical systems and bifurcation theory we have shown the emergence of spiking dynamics in our system. In our case, the emergence of excitability is related to the presence of a homoclinic bifurcation where the oscillatory state is destroyed (Sec.~\ref{sec:2}). Excitability mediated by homoclinic bifurcation is known as type-I and is characterized by the presence of a real threshold for the perturbations and by the divergence of the limit cycle period involved in the process as approaching its destruction \cite{izhikevich_neural_2012}. Excitability and spiking behavior are described in Sec.~\ref{sec:3}, where spike trajectory in a reduced phase space is compared with a perturbation below a threshold.  
Finally, we have also studied the implications that additive Gaussian noise may have on the spiking dynamics (Sec.~\ref{sec:4}). We show that a weak noise is able to excite the system triggering spikes randomly and irregularly. As the noise increases, the spike frequency increases and the inter-spike time reduces, leading to the appearance of regular oscillations for an optimal noise value. This process is known as coherence resonance \cite{pikovsky_coherence_1997}.  

This model predicts the presence of excitability in different platforms of photonic dimers that can find applications as integrated networks of nonlinear optical cells \cite{brunstein}. Also, spiking and neuron-like behaviors can find applications in brain-inspired softwares and hardwares. \cite{Feldmann2019, Xiang:20}

\section*{Acknowledgments}

This work was supported by the Fonds de la Recherche Scientifique - FNRS under grant No PDR.T.0104.19 and the European Research Council (ERC) under the European Union’s Horizon 2020 research and innovation program (grant agreement No 757800). F.L. acknowledges the support of the Fonds de la Recherche Scientifique-FNRS). P. P. -R acknowledges support from the European Union’s Horizon 2020 research and innovation program.
under the Marie Sklodowska-Curie grant agreement no. 101023717.

\bibliography{biblio}
\end{document}